\documentclass{pasj00}
\usepackage{color}
\usepackage{hangcaption}
\begin{document}
\SetRunningHead{\textsc{Uzawa} et al.}{A Large X-ray Flare from a Single Weak-lined T Tauri Star TWA-7 Detected with MAXI GSC}

\title{A Large X-ray Flare from a Single Weak-lined T Tauri Star TWA-7
Detected with MAXI GSC}


\author{Akiko \textsc{Uzawa},\altaffilmark{1} Yohko \textsc{Tsuboi},\altaffilmark{1} Mikio
\textsc{Morii},\altaffilmark{2} Kyohei
\textsc{Yamazaki},\altaffilmark{1} Nobuyuki \textsc{Kawai},\altaffilmark{2,4} Masaru
\textsc{Matsuoka},\altaffilmark{3,4} Satoshi \textsc{Nakahira},\altaffilmark{5} Motoko \textsc{Serino},\altaffilmark{4} Takanori \textsc{Matsumura},\altaffilmark{1} Tatehiro \textsc{Mihara},\altaffilmark{4} Hiroshi \textsc{Tomida},\altaffilmark{3} Yoshihiro \textsc{Ueda},\altaffilmark{8} Mutsumi \textsc{Sugizaki},\altaffilmark{4} Shiro \textsc{Ueno},\altaffilmark{3} 
Arata \textsc{Daikyuji},\altaffilmark{11} Ken \textsc{Ebisawa},\altaffilmark{10} Satoshi
\textsc{Eguchi},\altaffilmark{8} Kazuo \textsc{Hiroi},\altaffilmark{8} Masaki
\textsc{Ishikawa},\altaffilmark{12} Naoki \textsc{Isobe},\altaffilmark{8} Kazuyoshi
\textsc{Kawasaki},\altaffilmark{3} Masashi \textsc{Kimura},\altaffilmark{7} Hiroki
\textsc{Kitayama},\altaffilmark{7} Mitsuhiro \textsc{Kohama},\altaffilmark{3} Taro
\textsc{Kotani},\altaffilmark{5} Yujin E. \textsc{Nakagawa},\altaffilmark{4} Motoki
\textsc{Nakajima},\altaffilmark{9} Hitoshi \textsc{Negoro},\altaffilmark{6} Hiroshi
\textsc{Ozawa},\altaffilmark{6} Megumi \textsc{Shidatsu},\altaffilmark{8} Tetsuya \textsc{Sootome},\altaffilmark{4} Kousuke \textsc{Sugimori},\altaffilmark{2} 
Fumitoshi \textsc{Suwa},\altaffilmark{6} Hiroshi \textsc{Tsunemi},\altaffilmark{7} Ryuichi
\textsc{Usui},\altaffilmark{2} Takayuki \textsc{Yamamoto},\altaffilmark{4} Kazutaka
\textsc{Yamaoka},\altaffilmark{5} and Atsumasa \textsc{Yoshida},\altaffilmark{4,5} 
}

\altaffiltext{1}{Department of Physics, Faculty of Science and
Engineering, Chuo University, 1-13-27 Kasuga, Bunkyo-ku, Tokyo 112-8551,
Japan}
\email{akiko@phys.chuo-u.ac.jp}
\email{tsuboi@phys.chuo-u.ac.jp}
\altaffiltext{2}{Department of Physics, Tokyo Institute of Technology,
2-12-1 Ookayama, Meguro-ku, Tokyo 152-8551, Japan}
\altaffiltext{3}{ISS Science Project Office, Institute of Space and
Astronautical Science (ISAS), Japan Aerospace Exploration Agency (JAXA), 2-1-1 Sengen, Tsukuba, Ibaraki 305-8505, Japan}
\altaffiltext{4}{MAXI team, RIKEN, 2-1 Hirosawa, Wako, Saitama 351-0198, Japan}
\altaffiltext{5}{Department of Physics and Mathematics, Aoyama Gakuin
University,5-10-1 Fuchinobe, Chuo-ku, Sagamihara, Kanagawa 252-5258,
Japan}
\altaffiltext{6}{Department of Physics, Nihon University, 1-8-14
Kanda-Surugadai, Chiyoda-ku, Tokyo 101-8308, Japan}
\altaffiltext{7}{Department of Earth and Space Science, Osaka
University, 1-1 Machikaneyama, Toyonaka, Osaka 560-0043, Japan}
\altaffiltext{8}{Department of Astronomy, Kyoto University, Oiwake-cho, Sakyo-ku, Kyoto 606-8502, Japan}
\altaffiltext{9}{School of Dentistry at Matsudo, Nihon University,
2-870-1 Sakaecho-nishi, Matsudo, Chiba 101-8308, Japan}
\altaffiltext{10}{Department of Space Science Information Analysis,
Institute of Space and Astronautical Science (ISAS), Japan Aerospace Exploration Agency(JAXA), 3-1-1 Yoshino-dai, Chuo-ku, Sagamihara, Kanagawa 252-5210, Japan}
\altaffiltext{11}{Department of Applied Physics, University of Miyazaki,
1-1 Gakuen Kibanadai-nishi, Miyazaki, Miyazaki 889-2192, Japan}
\altaffiltext{12}{School of Physical Science, Space and Astronautical
Science, The graduate University for Advanced Studies (Sokendai), Yoshinodai 3-1-1, Chuo-ku, Sagamihara, Kanagawa 252-5210, Japan}

%


\maketitle
\begin{abstract}

We present a large X-ray flare from a nearby weak-lined T Tauri star TWA-7 detected with the Gas Slit Camera (GSC) on the Monitor of All-sky X-ray Image (MAXI).
The GSC captured X-ray flaring from TWA-7 with a flux of $3\times10^{-9}$ ergs cm$^{-2}$ s$^{-1}$ in 2--20 keV band during the scan transit starting at UT 2010-09-07 18:24:30.
The estimated X-ray luminosity at the scan in the energy band is 3$\times10^{32}$ ergs s$^{-1}$, 
indicating that the event is among the largest X-ray flares from
T Tauri stars.
Since MAXI GSC monitors a target only 
during a scan transit of about a minute per 92 min orbital cycle,
the luminosity at the flare peak might have been higher
than that detected.  
At the scan transit, we observed a high X-ray-to-bolometric luminosity ratio, log $L_{\rm X}$/$L_{\rm bol}$ = $-0.1^{+0.2}_{-0.3}$; i.e., the X-ray luminosity is comparable to the bolometric luminosity.
Since TWA-7 has neither an accreting disk nor a binary companion, the observed event implies that none of those are essential to generate such big flares
in T Tauri stars.
\end{abstract}

\section{Introduction}

TWA-7 (2MASS J10423011-3340162) is a weak-lined T Tauri star
first recognized with a follow-up observation of unidentified ROSAT
X-ray sources (Voges et al. 1999; Webb et al. 1999; Neuh\"auser et
al. 2000).  It is a member of the nearby association of young stars,
TW Hydrae Association (TWA; Kastner et al. 1997).  The distance to the
source is estimated to be 27$\pm$2 pc by Mamajek (2005) from the
moving cluster method (e.g., Atanasijevic 1971; de Bruijne 1999). We
use this value as the distance throughout this paper, although in the
past literature, $d = 55\pm16$ pc had been usually adopted from the
apparent extent of TWA and the mean Hipparcos distance derived from
four members of TWA (Neuh\"auser et al. 2000). TWA-7 has the spectral
type of M1 and youth indicators of high lithium abundance, X-ray
activity, and strong chromospheric activity (Webb et al. 1999;
Neuh\"auser et al. 2000).  The bolometric luminosity at 27 pc is
re-calculated to be log $L_{\rm bol}$/$L_{\odot} \sim -1.0$ from the
value obtained by Neuh\"auser et al. (2000) who assumed that $d = 55$ pc.
The excess emission detected with infrared (24 $\mu$m and 70 $\mu$m; Low et al. 2005)
and submillimeter bands (Webb 2000; Matthews et al. 2007) indicates the presence 
of a debris disk without ongoing accretion.
The mass of the disk is estimated to be 4.3 $M_{lunar}$, if the distance is 27 pc. 


Owing to the close proximity, the multiplicity of TWA-7 has been well
studied with imaging observations. Neuh\"auser et al. (2000) argued
the possibility of direct imaging detection of extra-solar planets
with ground-based telescopes, and raised the evidence for a possible
planetary companion to TWA-7 using HST NICMOS. Assuming the distance
of 55 pc, they estimated the mass of 3 $M_{jupiter}$ for the faint
object. A further search with the HST NICMOS revealed that the faint
possible companion is a background object from the inconsistency in
proper motions (Lowrance et al. 2005), but the above studies indicate
that any possible companion of TWA-7 must have a very low mass, possibly
planetary mass.

In the ROSAT All-Sky Survey catalog (Voges et al. 1999), TWA-7 is
named as 1RXS J104230.3-334014.
An X-ray flux of 3$\times10^{-12}$ ergs cm$^{-2}$ s$^{-1}$ is derived
in the ROSAT band ($\sim$0.1--2 keV) by Webb et al. (1999) and Stelzer
\& Neuh\"auser (2000). TWA-7 is also detected through the XMM-Newton
Slew Survey 
in 2010 January at 1.9 counts s$^{-1}$ (0.2--12 keV) and 1.6 counts s$^{-1}$ (0.2--2 keV) (Saxton et al. 2008). The hardness ratio obtained by dividing the count rate in
2--12 keV by that in 0.2--2 keV band is
0.19. If we assume a thin thermal spectrum ($apec$ model in XSPEC;
Smith et al. 2001) with metal abundance of 0.3 solar and negligible
absorbing column, a temperature of $kT = 2$ keV is consistent with 
the observed hardness ratio.
This model gives the flux of 4$\times10^{-12}$ ergs
cm$^{-2}$ s$^{-1}$ (0.2--2 keV), and 2$\times10^{-12}$ ergs cm$^{-2}$
s$^{-1}$ (2--12 keV). The flux in the 0.2--2 keV band is almost equal to
that obtained with ROSAT, and we can be confident that these fluxes
correspond to the quiescent level.

We detected a large X-ray flare from TWA-7 with the Gas Slit Camera (GSC) 
on the Monitor of All-sky X-ray Image (MAXI) 
deployed on the International Space Station (ATel \#2836: Morii et al. 2010).
The X-ray flux is three orders of magnitude
higher than that in the quiescent level, and the X-ray luminosity indicates 
this event is among the largest X-ray flares from T Tauri stars.

\section{Observations and Results}

The MAXI carries two scientific instruments: the Gas Slit Camera (GSC)
(Mihara et al. 2011) and the Solid State Camera (SSC) (Tomida et
al. 2010). Both the GSC and the SSC 
consist of two identical modules aimed at
different directions
(horizontal and zenithal direction). The instantaneous field of view
for one direction is 3$^{\circ}\times$160$^{\circ}$ (GSC) and
1.5$^{\circ}\times$90$^{\circ}$ (SSC), if all the cameras are
active. It means that they can cover 98\% (GSC) and 70\% 
(SSC) of the whole sky at the maximum every orbit, if there is no down time on the detectors.
The MAXI surveys the sky 16 times per day. The GSC consists of twelve
one-dimensional position-sensitive proportional counters operating in
the 2--30 keV range, while the SSC is composed of 32 X-ray CCD cameras
covering the energy range of 0.5--12 keV.  The GSC offers larger
effective area (5350 cm$^2$) than that of the SSC (200 cm$^2$). See
Matsuoka et al. (2009) for more details about the MAXI.  On 2010
September 7th, two GSC cameras, covering the direction of TWA-7,
caught an intense flare, while no data were obtained for the
same direction with SSC.

The transient X-ray emission was observed at the position of (R.A., Dec)
= (10d 43m 27s, $-$33h 40m 53s) (J2000) during the scan transit of 40 seconds starting at UT 2010-09-07 18:22:45.
The error region (90\% confidence level) can be estimated with a combination of the statistical error box and the calibration uncertainty on the alignment (0.2 degree at the 90\% limit; Sugizaki et al. 2011).
The statistical error box is a rectangle with the long side of 0.845 degree and the short side of 0.565 degree.
The corners are (R.A., Dec) = (10d 44m 22s, -34h 07m 49s), (10d 40m 55s, -33h 40m 54s), (10d 42m 32s, -33h 13m 44s), and (10d 45m 58s, -33h 40m 36s) (J2000). Within the error region, only TWA-7 was found in the ROSAT bright source catalog (Voges et al. 1999).
In order to test the significance of the transient event, we counted the numbers of events
in the 2--20 keV band in 18 circles with 1.5$^{\circ}$ radius around the transient event.
Here, the radius 1.5$^{\circ}$ corresponds to the HWHM (Half-Width Half Maximum) of the PSF (Point Spread Function) of MAXI GSC. We adopted the standard deviation of these 18 measurements
for the 1-sigma background fluctuation.
We extracted the counts in the same X-ray band in the circle with the same radius centered at the transient event, and subtracted the average of the background counts.
Then we confirmed that the transient event is detected with 12-$\sigma$ level, and
at the previous and the following scan transits (92 min before and after the detection), 
no significant X-ray emission was detected at more than 3-$\sigma$ level. 
We, hereafter, call each of the above transits Phase 0 (pre-flare), Phase 1 (flare), and Phase 2 (post-flare), respectively.

\begin{figure}[htbp]
\begin{center}
\includegraphics[width=80mm]{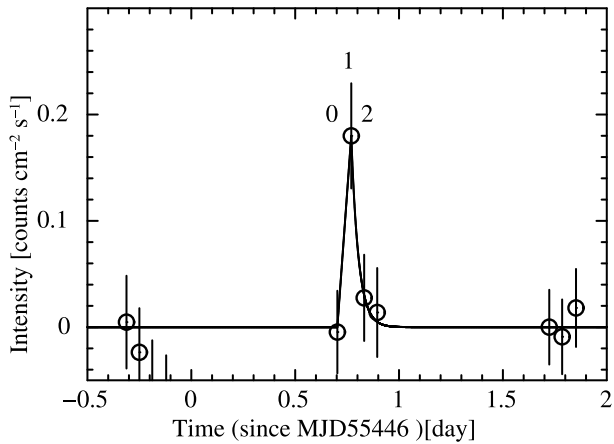}
\end{center}
\caption{GSC light curve of TWA-7 in the 2--20 keV band (the errors
 are 1-$\sigma$). The time span for one bin is about 1 minute which correspondents to one scan transit.
Each inserted number (0, 1, 2) is allocated to a scan transit in pre-flare, flare, and decay phases, respectively.
The solid
 line indicates the best-fit burst model (linear rise followed by an
 exponential decay). 
}\label{Fig1}
\end{figure}

Figure \ref{Fig1} shows the background-subtracted X-ray light curve in
the 2--20 keV band.
The time span for one bin is about 1 minute, which corresponds to one scan transit.
The data were extracted from an 
ellipse with 1.8$^{\circ}$ semi-major axis and 2.5$^{\circ}$ semi-minor axis, centered on TWA-7. This region was selected to maximize the S/N ratio.
The background was extracted from a circle with radius of 10$^{\circ}$
centered on TWA-7, by removing the source region. 
In Figure \ref{Fig1}, the error bar indicates 1-$\sigma$ error of the background-subtracted events, which is derived from $\sqrt[]{ \mathstrut S^{2} + B^{2}}$ (S is the counts in the source region and B is background counts which is expected to enter the source region).
We fitted the light curve with a burst model which shows a linear rise followed by an
exponential decay. The upper limit of the rising time is derived as 1.6 hours from the time-span between Phase 0 and 1,
and the $e$-folding time is derived to be
${\leq}$2.1 hours (90\% confidence range) by fitting from Phase 1 to the last bin in Figure 1.
These values are not inconsistent with the stellar flares reported in
the literature (e.g. Imanishi et al. 2003, Stelzer, Neuh\"auser, \& Hambaryan 2000, Pandey \& Singh 2008).

\begin{figure}[htbp]
\begin{center}
\includegraphics[width=80mm]{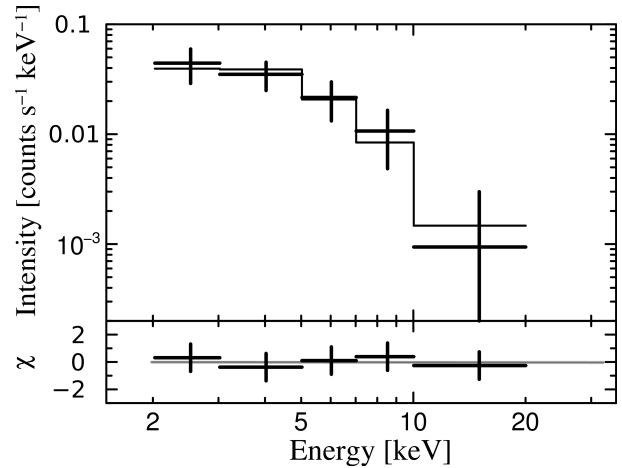}
\end{center}
\caption{GSC spectrum of TWA-7 at Phase 1 (the errors
 are 1-$\sigma$). The solid line shows the
 best-fit model ($apec$). Lower panel shows the residuals from the best-fit model.}\label{Fig2}
\end{figure}

Figure \ref{Fig2} shows the background-subtracted X-ray spectrum during the flare phase (Phase 1).
The source region and the background region are the same as those used to
make the light curve.
We fitted the spectrum with the optically-thin thermal plasma
model ($apec$ model) in which all the metal
abundances were fixed to 0.3 of solar values (Anders et al. 1989), 
which is generally obtained in various star forming regions (e.g. Imanishi et
al. 2001, Scelsi et al. 2007).
Since the interstellar absorption toward TWA-7 is negligible, 
we fixed the absorbing column to zero. The best-fit values are summarized in Table \ref{Tab2}, while the best-fit model is shown as the solid line in Figure \ref{Fig2}.
We obtained 3$\times10^{-9}$ ergs s$^{-1}$ cm$^{-2}$ for the flux in the 2--12 keV band. It is three orders of magnitude larger than the quiescent level ($\sim$2$\times10^{-12}$ ergs cm$^{-2}$ s$^{-1}$ see \S1).
The derived emission measure and X-ray luminosity in the 2--20 keV band are 2 (1--4) $\times$10$^{55}$ cm$^{-3}$ and 3 (2--4) $\times$10$^{32}$ ergs s$^{-1}$, respectively (errors are 90\% confidence range).
We also investigated the previous and following X-ray activity of this source using the entire MAXI GSC data record for about 1.5 years since the operation started in 2009 August 15,
but found no significant flux except for the time span that this flare occurred
on 2010 September 7.
The upper limit is 1$\times10^{-11}$ ergs s$^{-1}$ cm$^{-2}$ in the 4--10 keV band 
with a detection significance ($s_{\rm D}$) of 7 (Hiroi et al. 2011).
The detection limit of MAXI GSC is not sensitive enough to detect the quiescent X-ray flux of TWA-7 that one expects on the basis of previous measurements with different X-ray satellites.
One might suspect that the MAXI GSC band, which extends to 20 keV, can constrain the presence of a non-thermal component, but with the poor statistics, the spectrum does not provide evidence for the component.

\begin{table*}[htbp]
 \begin{center}
  \caption{Best-fit parameters to GSC spectrum of TWA--7 at Phase 1. }\label{Tab2}
  {\footnotesize
 \begin{tabular}{ccccccc} 
  \hline \hline
  $N_{\rm{H}}$ & $kT$ & $Z$ & $\int_{}^{} n_{e}n_{\rm H}dV$ & Flux$^{\dagger}$ & $L_{\rm X}^{\dagger}$ & $\chi^{2}_{\nu}$ (d.o.f) \\
  ($10^{22}\hspace{1ex} {\rm cm}^{-2}$) & (keV) & (solar abundance) & ($10^{55}\hspace{1ex} {\rm cm}^{-3}$) & ($10^{-9} {\rm ergs}\hspace{1ex} {\rm cm}^{-2}\hspace{1ex} {\rm s}^{-1}$) & ($10^{32} {\rm ergs\hspace{1ex} s}^{-1}$) & \\ \hline
  0 (fixed) & 6 & 0.3 (fixed) & 2 & 3 & 3 & 0.16 (3) \\
  & (3--27) & & (1--4) &(2--5) &(2--4) &  \\ \hline 
 \end{tabular}}
 \end{center}
  {\footnotesize
 \begin{tabular}{l}
  Note -- Error range refer to 90 \% confidence intervals for a single parameter.\\ $^{\dagger}$ : Flux and $L_{\rm X}$ are obtained in the 2--20 keV
  band. Distance is assumed to be 27 pc.\\
 \end{tabular}}
\end{table*}

\section{Discussion and Conclusion}

We have detected an X-ray flare from TWA-7 for the first time. The X-ray luminosity in the 2--20 keV band reached up to $3\times10^{32}$ ergs s$^{-1}$, which is relatively large in flares from T Tauri stars.
This luminosity, however, gives only the lower limit on the flare peak,
because MAXI GSC can monitor a target only once ($\sim$1 min) per 92 min orbital cycle.
During Phase 1, which would not be the real peak, 
log $L_{\rm X}$/$L_{\rm bol} = -0.1^{+0.2}_{-0.3}$ is obtained; i.e., the X-ray luminosity is comparable to the bolometric luminosity of 4$\times 10^{32}$ ergs s$^{-1}$ at the phase.
In various star-forming regions, the log $L_{\rm X}$/$L_{\rm bol}$ during flares from T Tauri stars range from $-2$ to $-4$ (e.g. $\rho$ Oph: Imanishi et al. 2001, Orion Trapezium at COUP project: Wolk et al. 2005). Even during the big flare from V773 Tau (Tsuboi et al. 1998), log $L_{\rm X}$/$L_{\rm bol} = -1.3$, which is still one order of magnitude smaller than that of TWA-7. Thus the most striking result obtained from the flare is the high X-ray-to-bolometric luminosity ratio.

What makes TWA-7 special? TWA-7 is known to have
neither a stellar mass companion nor an accretion disk (see \S1).
This 
implies that neither binarity (e.g. Getman et al. 2011) nor accretion (e.g. Kastner et al. 2002, Argiroffi et al. 2011), nor star-disk interaction (e.g. Hayashi et al. 1996, Shu et al. 1997, Montmerle et al. 2000) is essential to cause strong flares, but stellar magnetic activity alone can drive such energetic flare events.

An X-ray flare such as the one described herein with a large X-ray-to-bolometric ratio, log $L_{\rm X}$/$L_{\rm bol} \sim 0$, has not been reported before from a T Tauri star, but has been reported in other stellar categories. 
Pye \& McHardy (1983) and Rao \& Vahia (1987)
observed X-ray flares from RS CVn systems and dMe stars with the Sky Survey Instrument (SSI) on Ariel V, which scanned the sky once per $\sim$100 min with sensitivity in the 2--18 keV band. The observed log $L_{\rm X}$/$L_{\rm bol}$ ranges from $-2.9$ to $-0.1$.
With a trigger by the all sky monitor Swift BAT, Swift XRT also observed such a flare from the dMe star EV Lac with a log $L_{\rm X}$/$L_{\rm bol}$ of 0.49 in the 0.3--10 keV band.
Stellar flare events with log $L_{\rm X}$/$L_{\rm bol}$ of about 0 have only been found with all-sky surveys. This means that such giant flares are too rare to be detected with pointing observations.


This research has made use of the MAXI data\footnote{http://maxi.riken.jp/top/index.php}, provided by the RIKEN, JAXA, and MAXI teams. 
We thank Steven H. Pravdo for correcting English style and grammar.
This research was partially supported by the Ministry of Education, Culture, Sports, Science and Technology (MEXT), Grant-in-Aid No.19047001, 20041008, 20244015 , 20540230, 20540237, 21340043, 21740140, 22740120, and Global-COE from MEXT ``The Next Generation of Physics, Spun from Universality and Emergence'' and ``Nanoscience and Quantum Physics''.



\end{document}